\documentclass[final,5p,times,twocolumn]{elsarticle}
\usepackage{xspace,graphicx,color,pstricks,pst-plot,amssymb,amsmath,hyperref}
 
 \biboptions{sort&compress}
\newcommand{\jpsi}{\ensuremath{{\rm J}/\psi}\xspace}
\newcommand{\psip}{\ensuremath{\psi{\rm (2S)}}\xspace}
\newcommand{\chic}{\ensuremath{\chi_{c}}\xspace}

\newcommand{\chicOne}{\ensuremath{\chi_{c1}}\xspace}
\newcommand{\chicTwo}{\ensuremath{\chi_{c2}}\xspace}

\newcommand{\chiOne}{\ensuremath{\chi_{1}}\xspace}
\newcommand{\chiTwo}{\ensuremath{\chi_{2}}\xspace}
\newcommand{\oneSzero}{\ensuremath{{^1{\rm S}_0^{[8]}}}\xspace}
\newcommand{\threeSone}{\ensuremath{{^3{\rm S}_1^{[8]}}}\xspace}
\newcommand{\threePJ}{\ensuremath{{^3{\rm P}_J^{[8]}}}\xspace}

\newcommand{\pt}{\ensuremath{p_{\rm T}}\xspace}
\newcommand{\QQbar}{\ensuremath{Q \overline{Q}}\xspace}
\newcommand{\pTovM}{\ensuremath{p_{\rm T}/M}\xspace}
\newcommand{\chictwooverchicone}{\ensuremath{\chi_{c2}/\chi_{c1}}\xspace}

\newcommand{\chitwooverchione}{\ensuremath{\chi_{2}/\chi_{1}}\xspace}

\journal{Physics Letters B}

\begin{document}
\begin{frontmatter}

\title{Quarkonium production at the LHC:\\
a data-driven analysis of NRQCD's predictions}

\author[ist,lip]{Pietro Faccioli\corref{cor}}
\author[ist,cern]{Mariana Ara\'ujo}
\author[cern]{Valentin Kn\"unz}
\author[hephy]{Ilse Kr\"atschmer}
\author[cern]{Carlos Louren\c{c}o\corref{cor}} 
\author[ist,lip]{Jo\~ao Seixas}

\address[ist]{Physics Department, Instituto Superior T\'ecnico (IST), Lisbon, Portugal}
\address[lip]{Laborat\'orio de Instrumenta\c{c}\~ao e F\'{\i}sica Experimental de Part\'{\i}culas (LIP), Lisbon, Portugal}
\address[cern]{European Organization for Nuclear Research (CERN), Geneva, Switzerland}
\address[hephy]{Institute of High Energy Physics (HEPHY), Vienna, Austria}

\cortext[cor]{Corresponding authors: pietro.faccioli@cern.ch, carlos.lourenco@cern.ch}

\begin{abstract}
While non-relativistic QCD (NRQCD) foresees a variety of elementary quarkonium production mechanisms 
naturally leading to state-dependent kinematic patterns, 
the LHC cross sections and polarization measurements reveal a remarkably simple production scenario,
independent of the quantum numbers and masses of the quarkonia.
Surprisingly, NRQCD is able to accommodate the observed universal scenario,
through a series of conspiring cancellations smoothing out its otherwise variegated
hierarchy of mechanisms.
This seemingly unnatural solution implies that the $\chi_{c1}$ and $\chi_{c2}$ polarizations, 
not yet measured, are strong and opposite,
representing the only potential exception to a remarkably simple picture of quarkonium production.
The observation of a large difference between $\chi_{c2}$ and $\chi_{c1}$ polarizations,
which cannot be indirectly extracted from existing measurements because they mutually cancel 
each other in their contribution to the observed \jpsi\ production,
would be a smoking gun signal finally proving the multifaceted but mysteriously elusive 
structure of NRQCD.
On the other hand, the measurement of two similar, small polarizations 
will urge improved \mbox{P-wave} calculations, if not a substantial revision of the NRQCD hierarchies.
\end{abstract}

\end{frontmatter}

\sloppy

\section{Introduction}
\label{sec:intro}

Non-relativistic quantum chromodynamics (NRQCD)~\cite{bib:NRQCD} is generally considered as the best
theory approach to study quarkonium production.
It is important to compare its predictions to the experimental measurements 
recently made available by the LHC experiments, which already reach rather high
quarkonium transverse momentum, $\pt \approx 100$\,GeV, 
where the calculations are expected to be more reliable.
If discrepancies between the theory predictions and the measured patterns are found,
it is crucial to investigate if those differences stem from problems in the conceptual foundations
of the theory or from approximations and inaccuracies of the fixed-order perturbative calculations
available at present.

In a previous publication~\cite{bib:Paper1}, we considered the charmonium and bottomonium 
cross sections and polarizations measured at the LHC, in the mid-rapidity region, 
and brought to light their remarkably simple and universal patterns, 
when studied as a function of \pTovM, where $M$ is the quarkonium mass.
We concluded that all the measurements can be remarkably well described simply assuming 
universal production and decay properties for the ${^3{\rm S}_1}$ and ${^3{\rm P}_J}$ quarkonia.

In this Letter we investigate if 
the simplicity of the existing measurements can be reproduced by the NRQCD theory, 
despite its complexity and its foreseen hierarchy of elementary mechanisms,
as described in Section~3 of Ref.~\cite{bib:Paper1};
and, if so, in which specific conditions regarding the so-far unmeasured 
polarizations of the \chicOne\ and \chicTwo\ P-wave mesons.

As in Ref.~\cite{bib:Paper1}, we replace the calculated short-distance cross sections
(SDCs) with parametrized functions fitted to the experimental data, 
to be compared only a posteriori with the corresponding perturbative calculations.
We treat the direct \chicOne, \chicTwo\ and $\psi$ production cross sections as three
freely and independently varying kinematic functions.
However, in the absence of $\chi_c$ polarization measurements, 
we resort to calculated ingredients to fix them, 
according to two different scenarios, both inspired by NRQCD.

\section{Data-driven NRQCD scenarios}
\label{sec:NRQCDlikescenario}

\begin{figure*}[t]
\centering
\includegraphics[width=0.365\linewidth]{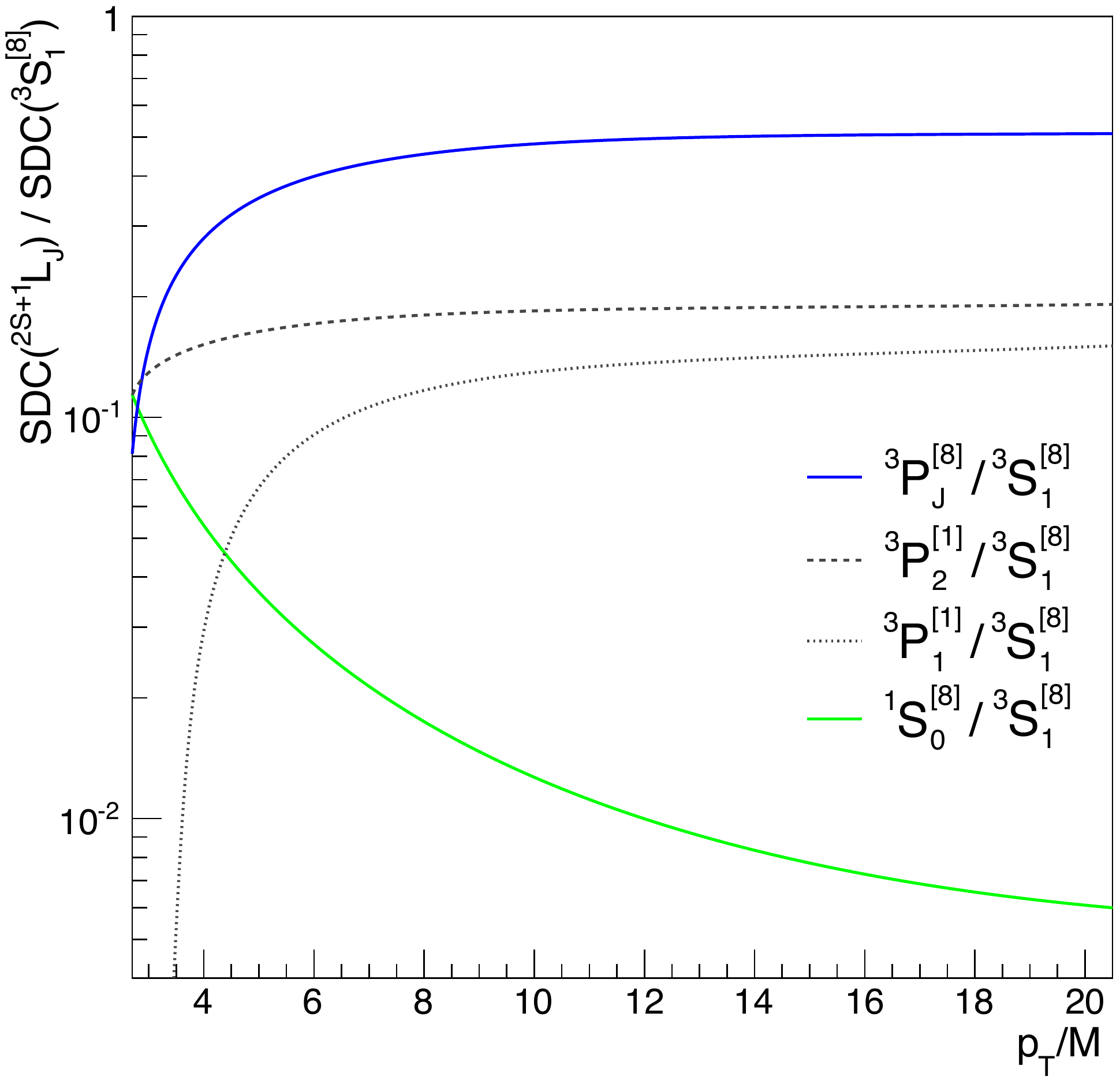}
\includegraphics[width=0.63\linewidth]{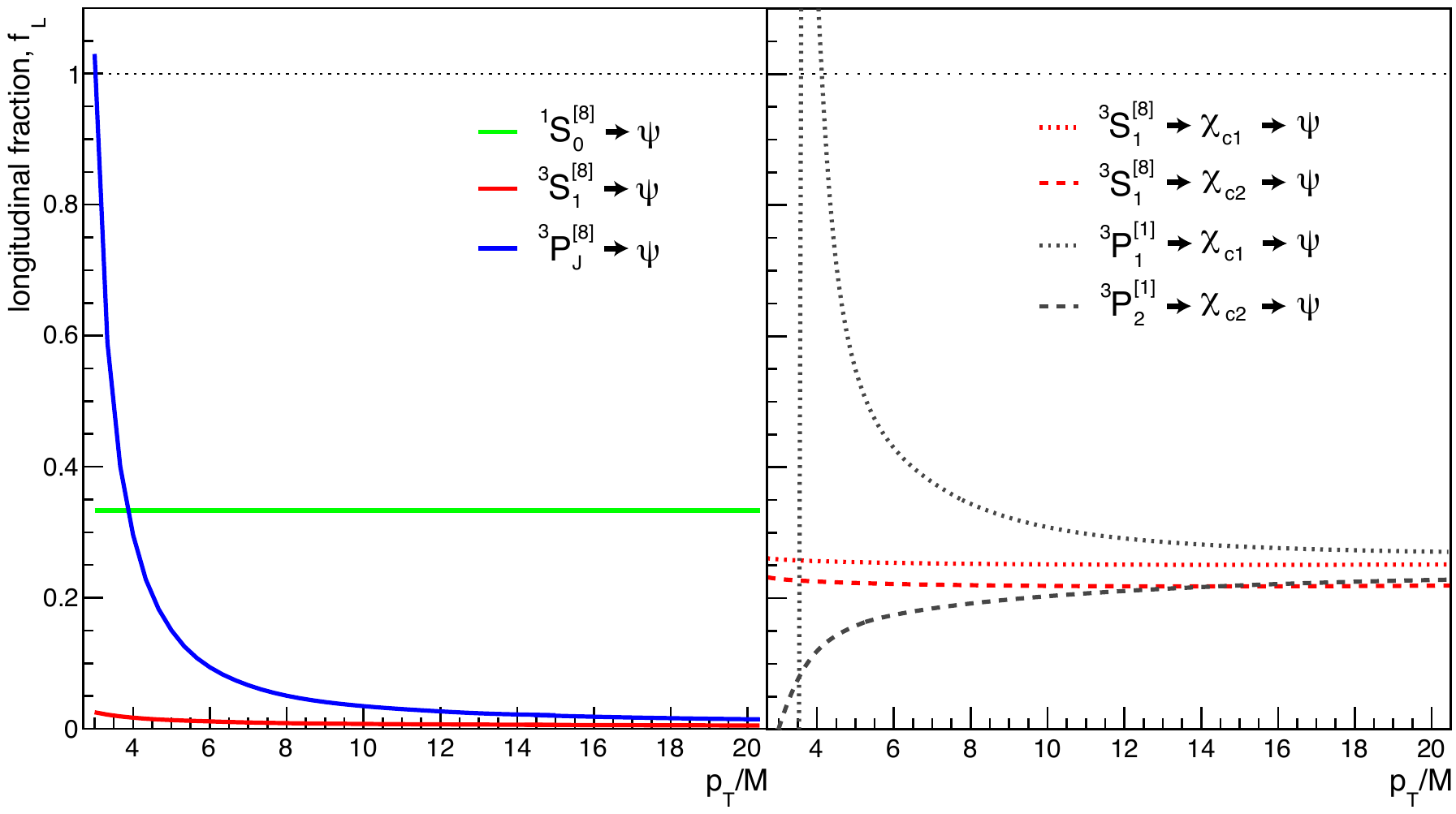}
\caption{Comparison of SDC trends~\cite{bib:Chao:2012iv, bib:Shao:2014fca, bib:Shao:2015vga} 
as a function of \pTovM:
SDC ratios with respect to the \threeSone\ reference (left);
longitudinal fractions for \mbox{S-wave} (``$\psi$") and \mbox{P-wave} (``$\chi$") production (right).
The \threePJ and ${^3{\rm P}_{1,2}^{[1]}}$ SDCs are multiplied by $m_c^2$, the squared charm-quark mass; they are negative and plotted with flipped signs.}
\label{fig:SDCratiosAndFL}
\end{figure*}

We have shown in Ref.~\cite{bib:Paper1} (Fig.~6-left) 
a surprisingly good agreement between the 
fully data-driven shape of the unpolarized component of the cross section
and that of the calculated NRQCD \oneSzero\ SDC. 
While the ``universal almost unpolarized" (UAU) scenario discussed there
reflects the simplest explanation of the data patterns, the conjecture
that $\chi$ and \jpsi\ productions are described by identical mixtures of the
same unpolarized and polarized processes finds no correspondence in the
hierarchies of the NRQCD factorization expansion, where, in particular, no
significant contribution of the unpolarized \oneSzero\ term is foreseen for
$\chi$ production. 
In this Letter we consider a scenario capable of accommodating the NRQCD process hierarchies, 
using two different assumptions for the $\chi$ sector. 

Direct \jpsi\ production remains completely flexible and can accommodate 
any production scenario of non-longitudinal polarization.
It is parametrized as a superposition of unpolarized and polarized processes,
characterized, respectively, by $\lambda_\vartheta=0$ and $\lambda_\vartheta=+1$,
where $\lambda_\vartheta$ is the polar anisotropy parameter of the $\psi$ dilepton decay.
In more concrete terms,
\begin{equation}
\label{eq:DirectXsect}
\sigma_{\rm dir}(\pTovM) = \sigma_{\rm dir}^{*} [ (1-f_{\rm p}^{*}) \, g_{\rm u}(\pTovM) + f_{\rm p}^{*} \, g_{\rm p}(\pTovM) ] \;,
\end{equation}
with $\sigma_{\rm dir}^{*}$ and $f_{\rm p}^{*}$ being, respectively, 
the total direct-production cross section and the fractional contribution of the polarized process, 
calculated at a reference point, set to $(p_{\rm T}/M)^{*} = 2$ in our study.
For $g_{\rm u}(\pTovM)$ and $g_{\rm p}(\pTovM)$, 
we take the empirical power-law functions defined in Eq.~5 of Ref.~\cite{bib:Paper1},
normalized at $(p_{\rm T}/M)^{*}$, 
determined by the corresponding shape parameters: $\beta_u$, $\beta_p$ 
and a common $\gamma$ (i.e., $\gamma_u \equiv \gamma_p$).
The \psip\ is identically parametrized, with the same shape parameters and $f_{\rm p}^{*}$.

Given the absence of $\chi$ polarization measurements, 
the modelling of the direct \chiOne\ and \chiTwo\ cross sections, instead, 
requires that we make a specific choice regarding these polarizations.
This choice represents the only model-dependent ingredient of the
analysis and the most substantial difference with respect to the UAU scenario,
where we imposed that both \chiOne\ and \chiTwo\ have identical process
composition as for \jpsi\ production.

To maintain full generality, we model the \chiOne\ and \chiTwo\ direct cross sections
as two individual and independent functions, 
defined by two additional power-law shape parameters, 
$\beta_{\chi_1}$ and $\beta_{\chi_2}$, 
while imposing the universal $\gamma$ 
(i.e., $\gamma_{\chi_1} \equiv \gamma_{\chi_2} \equiv \gamma$).
Therefore, we have a total of four independent and individually-observable 
cross-section contributions to quarkonium production: the unpolarized and polarized
$\psi$ terms, plus the \chiOne\ and \chiTwo\ ones. Their hypothetical NRQCD
counterparts are, respectively: \oneSzero, $\threeSone+\threePJ$,
$\threeSone+{^3{\rm P}_{1}^{[1]}}$ and $\threeSone+{^3{\rm P}_{2}^{[1]}}$.

We adopt two different $\chi$ polarization models. 
The first scenario, NRQCD\,1, deliberately in line with the criterion of simplicity guiding our analysis, 
is inspired by some recurring trends in the NRQCD perturbative calculations of \mbox{P-wave} components,
illustrated in Fig.~\ref{fig:SDCratiosAndFL}: with increasing \pTovM\ the
terms \threePJ, ${^3{\rm P}_{1}^{[1]}}$ and ${^3{\rm P}_{2}^{[1]}}$ become more
and more indistinguishable from the \threeSone\ term, in both cross section shapes and polarizations
(or longitudinal fractions, $f_{\rm L} = \sigma_{\rm L}/(\sigma_{\rm T}+\sigma_{\rm L})$, 
shown in the figure). 
This observation, also illustrated in Fig.~4 of Ref.~\cite{bib:Paper1},
suggests an hypothetical theory scenario where, as the perturbative calculations improve,
the three \mbox{P-wave} terms rectify their low-\pt\ behaviour,
tending to assimilate their shape to the one of their ``complementary'' \threeSone\ term.
This conjecture justifies our first modelling of the unknown \chiOne\ and \chiTwo\ polarizations: 
we assume them to be constant and equal to $1/5$ and $21/73$, respectively,
therefore similar to one another and to the measured \jpsi\ and \psip\
polarizations, in perfect coherence with the observed universal patterns.
These values are obtained with the relations
\begin{align}
\label{eq:chiPolarization}
\begin{split}
\lambda_\vartheta^{\chi_1} &= \lambda_\vartheta^{\jpsi} / (4 + \lambda_\vartheta^{\jpsi}) \; , \\
\lambda_\vartheta^{\chi_2} &= 21 \lambda_\vartheta^{\jpsi} / (60 + 13 \lambda_\vartheta^{\jpsi}) \; ,
\end{split}
\end{align}
when setting the decay anisotropy parameter of the \jpsi\ meson to
$\lambda_\vartheta^{\jpsi} = \lambda_\vartheta(\threeSone) = +1$.

The second scenario, NRQCD\,2, uses genuine NRQCD \chic\ polarization predictions,
obtained including ${^3{\rm P}_{1,2}^{[1]}}$ terms as currently 
calculated at NLO~\cite{bib:Chao:2012iv, bib:Shao:2014fca, bib:Shao:2015vga}.
In NRQCD, the $\chi$ polarizations and the \chitwooverchione\ cross-section
ratio are functions of one common parameter, equal for all $\chi_c$ states,
\begin{equation}
\label{eq:Kchi}
K_{\chi} = ({1}/{m_c^2}) \, \mathcal{L}_{\chi_{c0}}({^3{\rm P}_0^{[1]}}) / \mathcal{L}_{\chi_{c0}}({^3{\rm S}_1^{[8]}}) \; ,
\end{equation}
where $\mathcal{L}$ denotes the long distance matrix element (LDME) 
and $m_c^2$ is the mass of the charm quark, squared.
Because of heavy-quark spin-symmetry (HQSS), both the full $\chi_c$ production cross sections ($\sigma_J$, $J=0,1,2$) 
and the ``spin projections'' (spin-density matrix elements) $\sigma_J^{00}$\ldots $\sigma_J^{JJ}$ 
used to calculate $\lambda_\vartheta$ have the general form
\begin{equation}
\label{eq:chiXsectionsNRQCD}
\sigma_J \propto (2J+1) \left( m_c^2 \mathcal{S}({^3{\rm P}_J^{[1]}}) +  K_{\chi} \mathcal{S}({^3{\rm S}_1^{[8]}})  \right) \;,
\end{equation}
where, in the case of a spin-projected cross section, $\mathcal{S}$ 
represents the corresponding spin-projected SDC.
Using this equation to represent the numerator ($J=2$) and denominator ($J=1$) of the \chitwooverchione\ ratio,
we obtain $5/3$ when the singlet contributions vanish (\mbox{$K_{\chi}=0$}).
The spin-density matrix elements~\cite{bib:Chao:2012iv, bib:Shao:2014fca, bib:Shao:2015vga}
and $\lambda_\vartheta$ are related by
\begin{align}
\label{eq:chiLambdaVsSpinDensity}
\begin{split}
    \lambda_\vartheta^{\chi_1} & = ( \sigma_1^{00} - \sigma_1^{11} ) / ( \sigma_1^{00} + 3 \sigma_1^{11} ) \; , \\
    \lambda_\vartheta^{\chi_2} & = ( -3 \sigma_2^{00} - 3 \sigma_2^{11} + 6 \sigma_2^{22} ) / ( 5 \sigma_2^{00} + 9 \sigma_2^{11} + 6 \sigma_2^{22} ) \; ,
\end{split}
\end{align}
where each of the spin-density matrix elements is a linear combination of the 
${^3{\rm P}_J^{[1]}}$ and ${^3{\rm S}_1^{[8]}}$ terms, according to Eq.~\ref{eq:chiXsectionsNRQCD}. 
This induces the $\lambda_\vartheta$ dependence on $K_{\chi}$.
It is important to note that, for what concerns the $\chi$ polarization parameters 
$\lambda_\vartheta(\chi_{1,2})$, 
we always refer to the corresponding dilepton decay distributions of \jpsi\ from $\chi$. 
These are, in fact, the distributions that the experiments will measure directly 
and the ones needed to constrain the polarization of direct \jpsi\ production through a global analysis. 
Moreover, the dilepton decay parameters are identical to the $\chi \to \jpsi\ \gamma$ parameters, 
with the important advantage of being insensitive to the 
uncertain contributions of higher-order photon multipoles~\cite{bib:chiPol}.

\begin{figure}[t!]
\centering
\includegraphics[width=0.825\linewidth]{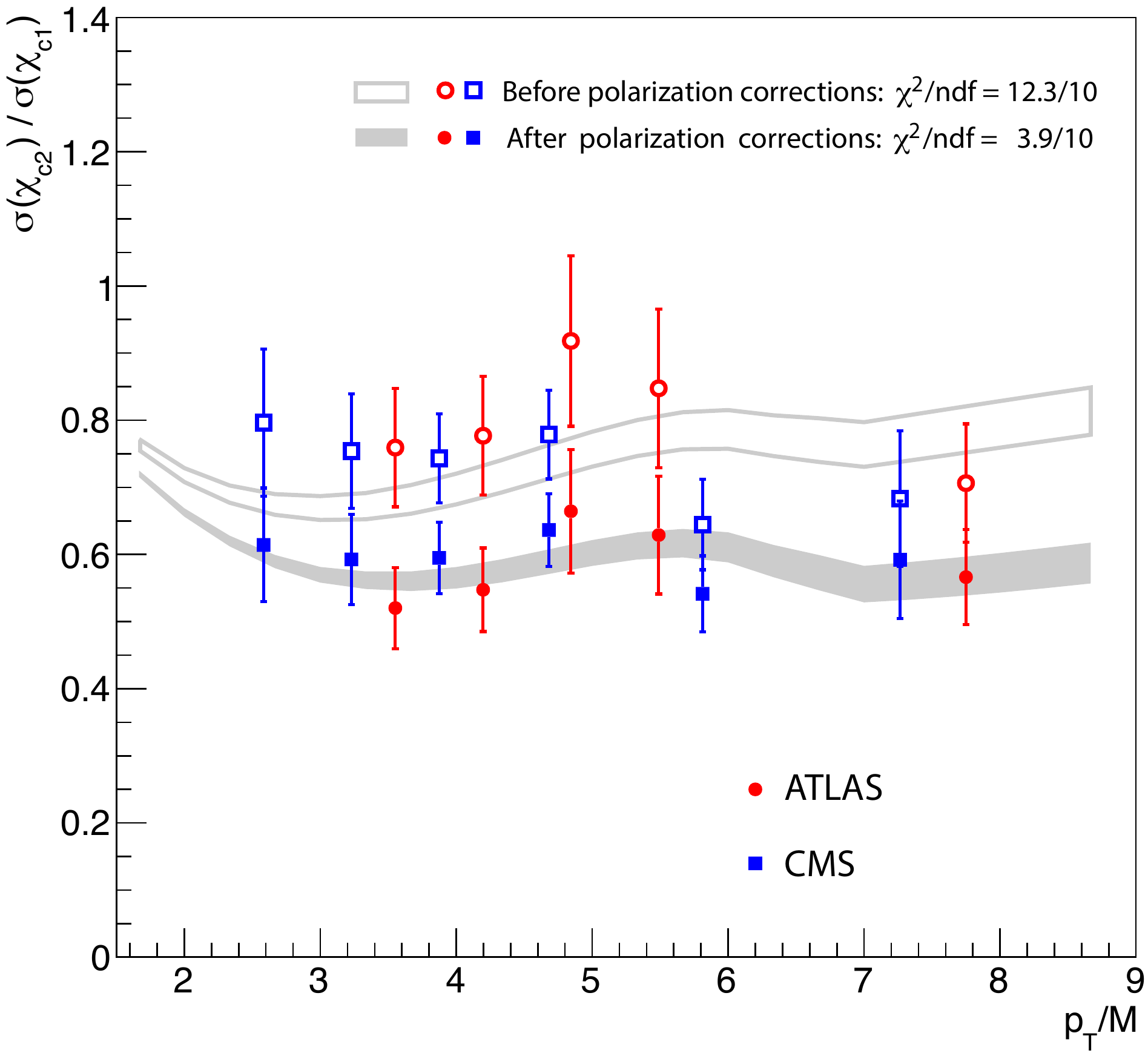}
\caption{Fit of the \chictwooverchicone\ ratio measured 
by ATLAS~\cite{bib:ATLASchic} 
and CMS~\cite{bib:CMSchic}, 
before and after properly accounting for the dependence of the detection acceptances on the
(simultaneously calculated) \chiOne\ and \chiTwo\ polarizations.
The filled markers are the data points acceptance-corrected according to 
the best-fit polarizations (shown in Fig.~\ref{fig:chiPolarizations}).
The width of the best-fit band reflects the $K_{\chi}$ uncertainty.}
\label{fig:chiRatioFitNRQCD}
\end{figure}

\begin{figure}[t!]
\centering
\includegraphics[width=0.85\linewidth]{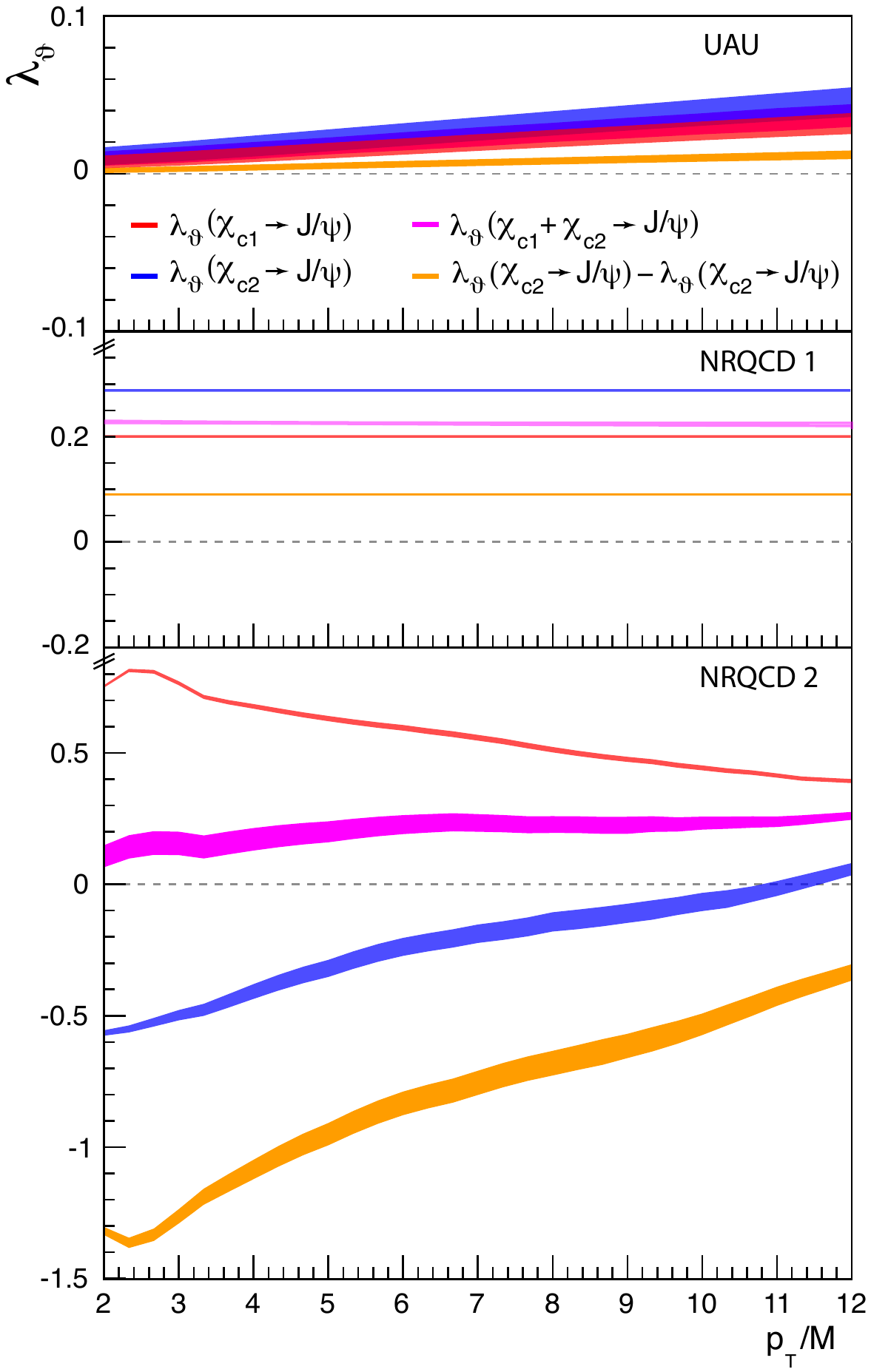}
\caption{\chicOne and \chicTwo polarizations as \emph{fitted} in the UAU scenario (top), 
with bands reflecting correlated parameter variations,
and as \emph{assumed} in the NRQCD\,1 (middle) and NRQCD\,2 (bottom) scenarios.
We also show the difference and the weighted sum 
(using the \jpsi\ feed-down contributions as weights)
of the two polarization parameters.
The NRQCD\,1 constant lines reflect the hypothesis of ${^3{\rm P}_{1,2}^{[1]}}$ polarizations 
assimilated to the \threeSone polarizations,
while the NRQCD\,2 curves are weighted sums of the \threeSone and ${^3{\rm P}_{1,2}^{[1]}}$ polarizations, 
with weights resulting from fitting the \chitwooverchione\ data (Fig.~\ref{fig:chiRatioFitNRQCD}).}
\label{fig:chiPolarizations}
\end{figure}

Leading-power fragmentation corrections have not yet been calculated for the 
${^3{\rm P}_{1,2}^{[1]}}$ singlets. Therefore, for consistency, 
we do not use, in Eq.~\ref{eq:chiXsectionsNRQCD},
the available \threeSone\ corrections~\cite{bib:BodwinCorrections}
in the case of $\chi$ production.

We can determine $K_{\chi}$ from the measured \chitwooverchione\ ratio and 
then calculate the corresponding NRQCD polarization predictions. 
However, we must take into account that the \chitwooverchione\ ratio measurements 
strongly depend on the \chiOne\ and \chiTwo\ polarizations assumed for the corrections
of the detector's acceptance (phase space coverage). 
Therefore, for every $K_{\chi}$ value considered, 
we calculate the \chiOne\ and \chiTwo\ polarization predictions and 
then correct the published ratio by the corresponding acceptance ratio. 
The corrected measurement is then compared (with statistical and systematic uncertainties, 
but without polarization uncertainties) with the prediction (for that $K_{\chi}$ value), 
to calculate the corresponding $\chi^2$ value.

\begin{figure*}[t!]
\centering
\includegraphics[width=0.75\linewidth]{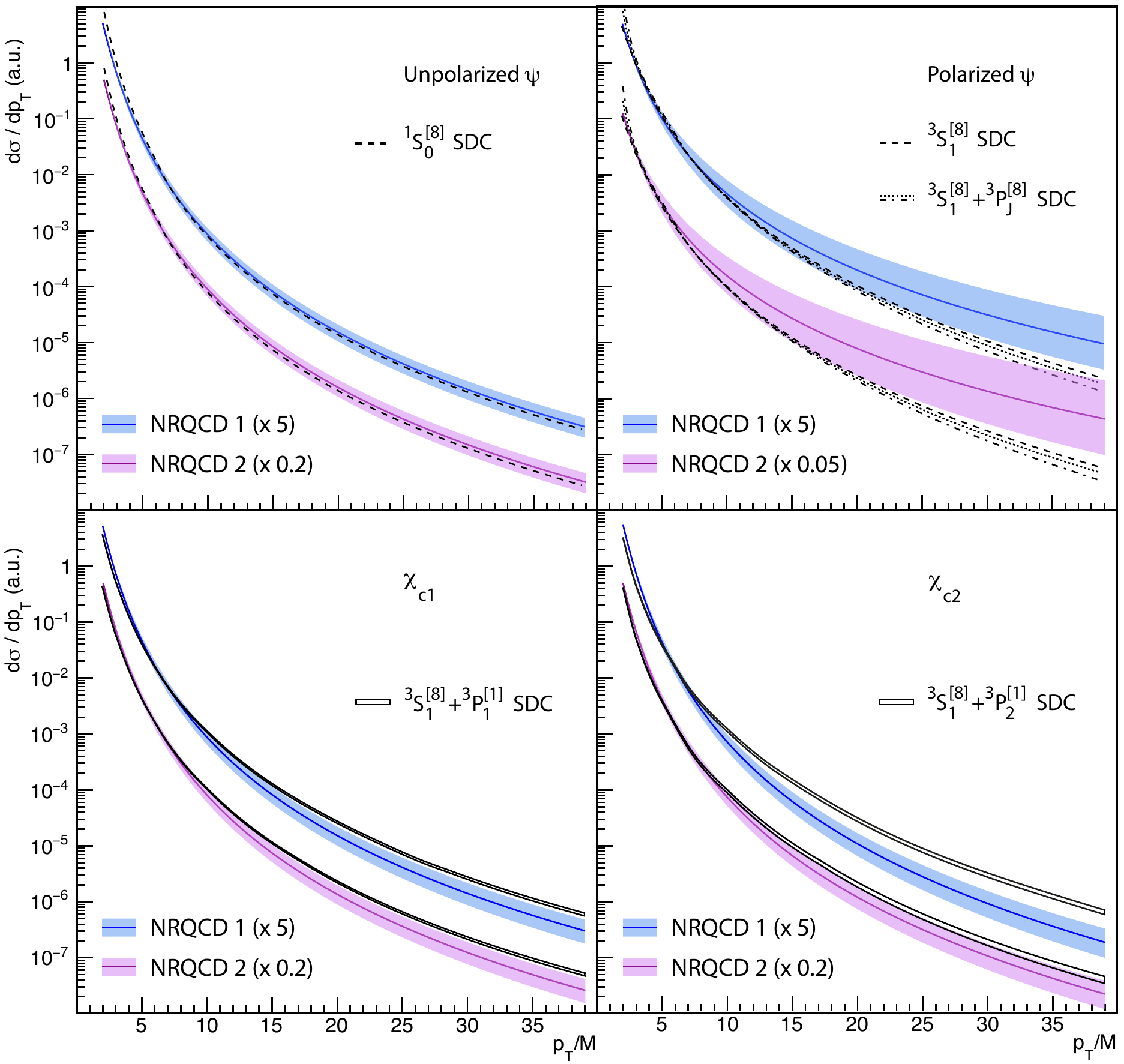}
\caption{Cross section contributions (normalized to unity at $\pTovM=2$) 
in the NRQCD\,1 and NRQCD\,2 scenarios,
with uncertainty bands reflecting correlated variations in the fit parameters.
The SDC combinations, arbitrarily normalized, are also shown
(with band widths reflecting the $K_\chi$ values given in the text).}
\label{fig:pTovMdistrNRQCD12}
\end{figure*}

Figure~\ref{fig:chiRatioFitNRQCD} shows how the \chitwooverchione\ ratio changes and the theoretical fit improves
when we use the ``proper'' NRQCD polarization conjecture instead of the default unpolarized scenario 
that the experiments assume to report the central values of the measurements. 
The result of the fit is $K_{\chi} = 4.60 \pm 0.06$, 
a value only marginally consistent with the (much more uncertain) value assumed in 
Ref.~\cite{bib:Shao:2014fca}
for the ${^3{\rm P}_J^{[1]}} + {^3{\rm S}_1^{[8]}}$ mixing parameter 
($1/K_{\chi} = 0.27 \pm 0.06$, i.e.\ $K_{\chi} = 3.7 ^{+1.0}_{-0.7}$ in our definition). 
Their result reflects two incorrect ingredients: 
the use of the unpolarized acceptance scenario 
and the interpretation of the entire spectrum of polarization hypotheses 
as part of the experimental uncertainty.

The polarization predictions obtained with our $K_{\chi}$ value, 
and used as input to the NRQCD\,2 scenario, are shown in Fig.~\ref{fig:chiPolarizations}.
As \pTovM\ decreases, they tend to the extreme physical values 
$\lambda_\vartheta = +1$ and $-3/5$ for the \chiOne\ and \chiTwo, respectively.
These maxima, while representing very different decay distributions, 
actually correspond to pure $J_z = 0$ alignment configurations 
of the two $\chi$ states, along an axis coinciding, in that limit, with the direction of the colliding partons.
Figure~\ref{fig:chiPolarizations} also shows the constant polarizations
used as inputs in the NRQCD\,1 scenario 
and the predictions resulting from the UAU fit~\cite{bib:Paper1}.

The NRQCD\,1 and NRQCD\,2 fits use the same data inputs as the UAU fit.
Both best-fit results are very similar to those seen in Fig.~5 of Ref.~\cite{bib:Paper1}:
the three scenarios describe the data very well, with almost identical minimum $\chi^2$ values.
The resulting shape parameters are, for the NRQCD\,1 and NRQCD\,2 fits, respectively:
\begin{align}
\label{eq:beta_gamma}
\begin{split}
& \gamma = 0.74 \pm 0.20 \; {\rm and} \; 0.73 \pm 0.19 \; , \\
& \beta_{\rm u} = 3.430 \pm 0.046 \; {\rm and} \; 3.421 \pm 0.045 \; , \\
& \beta_{\rm p} = 2.78 \pm 0.22 \; {\rm and} \; 2.67 \pm 0.29 \; , \\
& \beta_{\chi_1} = 3.440 \pm 0.082 \; {\rm and} \; 3.458 \pm 0.082 \; , \\
& \beta_{\chi_2} = 3.54 \pm 0.10 \; {\rm and} \; 3.49 \pm 0.10 \; .
\end{split}
\end{align}
The corresponding polarized fractions, at $(p_{\rm T}/M)^{*} = 2$, are
$f_{\rm p}^{*} = (1.8 \pm 1.5)\%$ and $(0.9 \pm 1.2)\%$.

The shapes of the cross-section terms obtained with the NRQCD\,1 and NRQCD\,2 fits
are shown in Fig.~\ref{fig:pTovMdistrNRQCD12},
the data-driven bands being compared with the corresponding NRQCD SDC combinations.
The unpolarized and polarized $\psi$ panels show, respectively, the \oneSzero\ and \threeSone\ SDCs,
calculated at next-to-leading order (NLO) with fragmentation contributions~\cite{bib:BodwinCorrections}.
The polarized panel also shows two curves representing a possible \threePJ\ contribution added 
to the \threeSone\ term, assuming 
$K_{\psi} = (1/{m_c^2}) \;
\mathcal{L}_{\jpsi}({^3{\rm P}_0^{[8]}}) \,/\, \mathcal{L}_{\jpsi}({^3{\rm S}_1^{[8]}}) = 0.1$ and 0.2.
Higher \threePJ\ contributions lead to steeper curves, further departing from the fit result, 
which prefers the \threeSone\ term alone. 
In fact, given the slight increase of the measured polarizations with increasing \pTovM, 
the data prefer a flatter polarized term than any $\threeSone + \threePJ$ combination.
Nevertheless, a residual \threePJ\ contribution cannot be excluded,
given the current experimental uncertainty; more precise polarization data, especially at high \pTovM, 
will make the comparison significantly more stringent.

\begin{figure}[t!]
\centering
\includegraphics[width=0.82\linewidth]{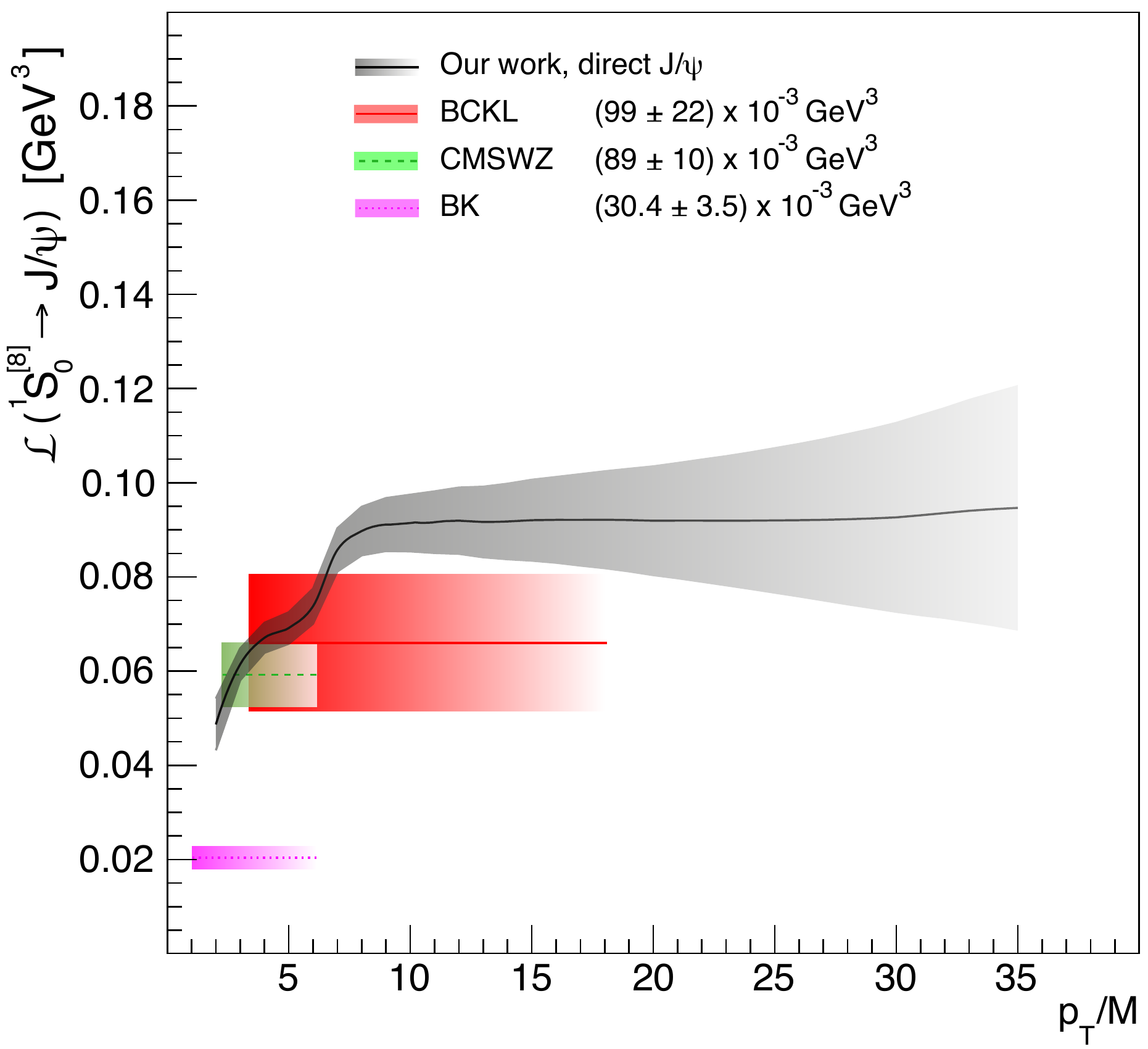}
\caption{\oneSzero LDME for direct \jpsi\ production in the NRQCD\,2 scenario (grey band),
compared to results obtained in other analyses: BK~\cite{bib:BK:PRD2011},
CMSWZ~\cite{bib:Chao:2012iv} and BCKL~\cite{bib:BodwinCorrections}.}
\label{fig:LDME1S08vspTovM}
\end{figure}

Making specific hypotheses about the physical nature of the unpolarized and polarized contributions
allows us to extract LDME values from the comparison with the calculated SDCs.
For example, identifying the unpolarized cross section with the \oneSzero\ term, 
the corresponding LDME for a given state is
\begin{equation}
\label{eq:LDMEcalc}
\mathcal{L}({^1{\rm S}_{0}^{[8]}}) =  ( 1-f_{\rm p} ) \,\sigma_{\rm dir} \,/\, \mathcal{S}({^1{\rm S}_{0}^{[8]}}) \; .
\end{equation}
Using the SDCs calculated at NLO complemented with fragmentation contributions,
for the production of a \QQbar\ pair with 3\,GeV of rest energy, 
the fit results for the \emph{direct} \jpsi\ cross section in the NRQCD\,2 scenario 
lead to the \oneSzero LDME represented by the grey band in Fig.~\ref{fig:LDME1S08vspTovM}. 
The width of the band only represents the experimental uncertainty; 
it does not reflect possible changes of the calculations due to scale dependencies 
(mostly a normalization shift) or to potential higher-order improvements. 

As evident in the figure and implied by Eq.~\ref{eq:LDMEcalc}, 
where $f_{\rm p}$, $\sigma_{\rm dir}$ and $\mathcal{S}({^1{\rm S}_{0}^{[8]}})$ are functions of \pTovM, 
the (supposedly constant) coefficient $\mathcal{L}({^1{\rm S}_{0}^{[8]}})$ becomes, 
technically, a function of \pTovM. 
This illustrates how the universality of the LDMEs, 
and hence the validity of the factorization hypothesis itself, 
are in reality only approximations relying on the goodness of the perturbative SDC calculations at a given order. 
The extracted LDME increases by a factor of 2 in the low-\pTovM region, 
presumably more affected by calculation uncertainties, and then stabilizes. 
Also for this reason, the LDME, when expressed by a constant number, 
necessarily depends on the method used for theory-data comparison. 
Figure~\ref{fig:LDME1S08vspTovM} also shows the $\mathcal{L}({^1{\rm S}_{0}^{[8]}})$ results 
obtained in three different analyses of charmonium production,
CMSWZ~\cite{bib:Chao:2012iv}, BCKL~\cite{bib:BodwinCorrections} and BK~\cite{bib:BK:PRD2011}, 
represented as uncertainty bands illustrating the \pTovM ranges of the used data. 
Since these results were extracted from prompt-\jpsi\ cross sections, 
neglecting \psip\ and $\chi_c$ feed-down, we rescaled them by $2/3$, 
the approximate ratio of direct-over-prompt \jpsi\ yields
(a practically \pTovM-independent ratio, given the almost universal \pTovM scaling of the cross sections). 

Remarkably, the three results follow a trend seemingly determined by the lower \pTovM limit 
chosen for the input data, 
which also reflects the highest event population and most constraining data points. 
That trend is well described by our curve, implicitly showing that all results are conceptually compatible,
while clearly revealing that the specific numerical results are somewhat arbitrary and mostly determined by 
the choice of the starting \pTovM value (or, in our case, by the normalization point in Eq.~\ref{eq:LDMEcalc}).
Our analysis method keeps data and theory disentangled, 
by using the data to obtain the best possible experimental determinations of global observables,
which are consecutively compared to theory. 
Such comparisons can, thus, be done as a function of kinematics, 
which is not possible when the theory calculations are intrinsically embedded in the fits. 

From the results of the NRQCD\,1 and NRQCD\,2 fits, 
where the \chicOne\ and \chicTwo\ cross section shapes are freely adjusted to the data 
with no imposed links to other contributions, 
we can appreciate the significance of the experimental indication towards identical \pTovM\ distributions 
for S- and \mbox{P-wave} states, which in the UAU scenario was a postulate. 
Indeed, the \chic\ shape functions are very similar to the unpolarized term dominating $\psi$ production,
as quantified by the compatibility of the $\beta_{\rm u}$, $\beta_{\chi_1}$ and $\beta_{\chi_2}$ values.
This also seems to contradict a priori expectations from NRQCD, 
where \chiOne\ and \chiTwo\ production should receive a major contribution from the \threeSone\ term, 
while the much flatter unpolarized \oneSzero\ term is strongly suppressed for these states. 
Although this picture could be somewhat modified by the \mbox{P-wave} singlet contributions 
(which, being negative, cannot be dominant), 
it is rather straightforward, in NRQCD, to expect flatter \pTovM\ dependencies for $\chi_c$
than for $\psi$ production. 
To understand the origin of this strong data indication, 
which might seem surprising given that the $\chi_c$ cross sections are measured at relatively low \pt\ 
and with comparatively low precision, 
we repeated the NRQCD\,1 fit keeping only one experimental point for each of the two \chic\ cross sections 
(chosen in the middle of the measured range), 
letting these measurements constrain the feed-down fractions at that point but not the \pTovM\ shapes. 
The results for the \chic\ cross section shapes remain fairly consistent with (and even steeper than) 
the unpolarized $\psi$ results:
$\beta_{\rm u} = 3.37 \pm 0.05$, 
$\beta_{\rm p} = 2.85 \pm 0.25$, 
$\beta_{\chi_1} = 3.57 \pm 0.10$, 
$\beta_{\chi_2} = 3.68 \pm 0.13$. 
This test reveals that what prevents the $\chi_c$ contribution to \jpsi\ production from having a 
flatter \pTovM\ dependence 
is the strong similarity between the precisely-measured \jpsi\ and \psip cross section shapes.

\begin{figure*}[t!]
\centering
\includegraphics[width=0.87\linewidth]{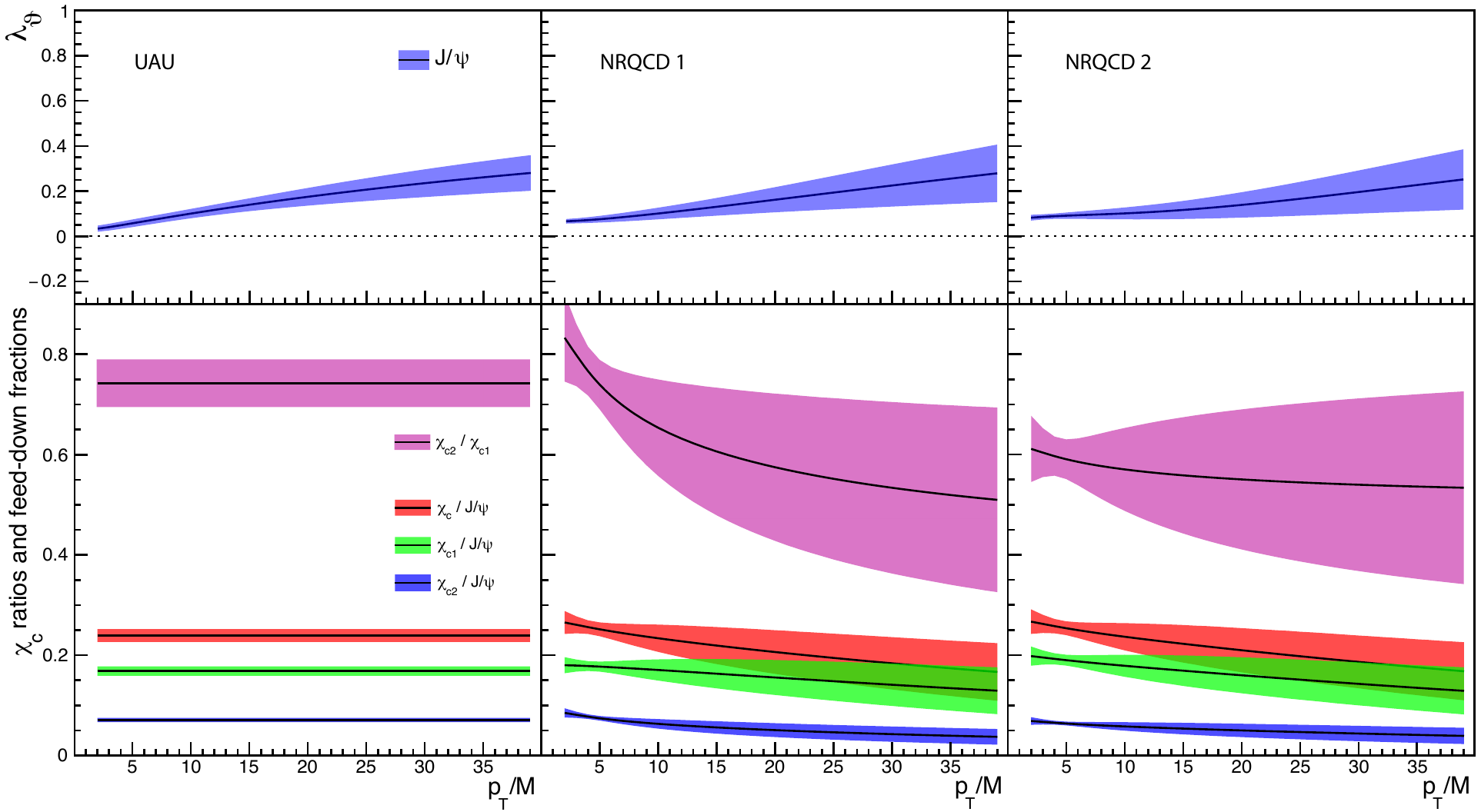}
\caption{The \jpsi\ polarization (top) and the \chictwooverchicone\ ratio and the
\jpsi\ feed-down fractions from \chicOne, \chicTwo and total-\chic (bottom),
in the UAU (left), NRQCD\,1 (middle) and NRQCD\,2 (right) scenarios.
The uncertainty bands reflect correlated variations in the fit parameters.}
\label{fig:otherPredictions}
\end{figure*}

The NRQCD\,1 \chicOne\ and \chicTwo\ $\threeSone + {^3{\rm P}_{1,2}^{[1]}}$ SDC combinations 
shown in Fig.~\ref{fig:pTovMdistrNRQCD12} were calculated with 
Eq.~\ref{eq:chiXsectionsNRQCD}, using $K_{\chi} = 4.13 \pm 0.09$,
the uncertainty leading to the difference between the two solid lines.
This $K_{\chi}$ value results from the fit of the \chictwooverchicone\ ratio,
correcting the detection acceptances using, this time, the NRQCD\,1 $\chi$ polarization scenario;
the best-fit curve is almost identical to the one shown in Fig.~\ref{fig:chiRatioFitNRQCD} 
for the unpolarized acceptance scenario.
We remind that no complete fragmentation corrections have been calculated for $\chi$ production 
and the available \threeSone corrections have not been used in this case.
The comparison seems to confirm the a priori expectation that NRQCD has some difficulties 
in reproducing \chic\ cross sections as steep as the \oneSzero-dominated \jpsi\ one, 
given the much flatter \threeSone SDC. 
Nevertheless, the negative ${^3{\rm P}_{1,2}^{[1]}}$ terms contribute significantly 
in the direction of a better compatibility with the data. 
It will be interesting to see if this trend will be perfected by perturbative calculations including fragmentation corrections.

Interestingly, the data-theory comparison improves further in the NRQCD\,2 scenario. 
The $m_c^2 \, \mathcal{S}({^3{\rm P}_J^{[1]}}) +  K_{\chi} \, \mathcal{S}({^3{\rm S}_1^{[8]}})$
combinations, with $K_{\chi}$ obtained from the \chitwooverchione\ fit assuming 
the ``full'' NRQCD $\chi$ polarization predictions with singlet terms, 
approach better the experimental data, as seen in the $\chi_{c2}$ cross section 
(Fig.~\ref{fig:pTovMdistrNRQCD12}, bottom-right) 
and in the \chitwooverchione\ fit itself (Fig.~\ref{fig:chiRatioFitNRQCD}).

We have seen that the \chic\ polarization hypothesis, 
sole model input to the otherwise fully data-driven scenarios we considered, 
does not have a crucial influence on the fit results. 
On the other hand, the NRQCD\,2 option brings the seemingly complex theoretical framework 
closer to a good description of what, at first sight, is an almost universal production picture.
Almost ironically, however, the NRQCD\,2 polarization hypothesis would represent, 
if confirmed experimentally, the only case of strong quarkonium polarizations observed by high-\pt\ experiments 
and the only evidence for a diversity in the production mechanisms of different states.

Another surprising coincidence is that the sum of the \chicOne\ and \chicTwo\ polarizations, 
weighted by their contributions to the \jpsi\ polarization (Fig.~\ref{fig:chiPolarizations}, pink band), 
is equally small and flat in the NRQCD\,2 scenario, as in the NRQCD\,1 option
(justifying that both models give very similar fit results). 
This fortuity prevents that the existence of such strong polarizations is
observable in the indirect \jpsi\ production. 
Even precise measurements of the difference between the \jpsi\ and \psip\ polarizations, 
or \chic\ measurements not resolving the $J=1$ and 2 states because of poor mass resolution, 
will not provide enlightnening information on the possible individual peculiarities of the \mbox{P-wave} polarizations.

Given these considerations, we argue that \chicOne\ and \chicTwo\ polarizations, 
measured individually, represent a distinctive experimental signature of NRQCD.
Fortunately, the almost maximal difference in their predicted values, 
when singlet contributions are taken into account, 
facilitates a conclusive experimental test. 
In particular, the observable $\lambda_\vartheta(\chi_2) - \lambda_\vartheta(\chi_1)$, 
also shown in Fig.~\ref{fig:chiPolarizations} (orange band), 
is almost immune to systematic uncertainties, 
like acceptance and efficiency determinations, 
and can be measured with maximal significance and accuracy.

Figure~\ref{fig:otherPredictions} shows how the
\jpsi\ polarization, \chictwooverchicone\ ratio and 
\jpsi\ feed-down fractions from \chicOne, \chicTwo and total-\chic
extrapolate from the current experimental trends to higher \pTovM, 
in the $\chi$ production scenarios we discussed.
These are the observables that, together with the $\chi$ polarizations, 
are in more pressing need of improved measurements by the LHC experiments.

\section{Discussion and conclusions}
\label{sec:conclusions}

The results presented in this Letter should be seen as a complement to the study shown in our previous
publication~\cite{bib:Paper1}, which also provides an extensive motivation and more details on the data used
and on the analysis methodology.
That paper is devoted to a phenomenological study, 
exclusively based on empirical functions derived directly from the measured patterns,
complemented by one data-inspired hypothesis:
all charmonia are born from a universal production mechanism (UAU), 
independently of the mass and quantum numbers of the state.

In the present Letter we went beyond the full universality hypothesis, 
allowing differences between the $\chi_c$ production mechanisms and those of
the \jpsi\ and \psip\ S-wave mesons.
We considered two different models of $\chi_c$ production, NRQCD\,1 and NRQCD\,2,
both possibly accommodating the NRQCD hierarchies of processes, 
with two degrees of complexity in the assumed $\chi_c$ polarization ingredients.
It is important to emphasise that reasonably precise $\chi_c$ polarization measurements,
once available, will free the procedure from the necessity of model assumptions,
presently imposed by the incomplete picture of quarkonium production provided by the
LHC experiments.

The two main questions addressed in this work are:
1)~how different and experimentally recognizable are the $\chi_c$ production mechanisms 
from those of the \jpsi\ and \psip\ mesons?
2)~and how important will be, for the understanding of quarkonium production,
new or improved $\chi_c$ measurements?

A first interesting point to mention is that, according to the data, 
the differential cross sections of the $\chi_c$ mesons,
as a function of \pTovM, cannot have a significantly different slope with respect to those of the
S-wave mesons, given the fact that the measured \jpsi\ and \psip\ spectra are almost identical
and knowing that around 25\% of the \jpsi\ yield results from $\chi_c$ feed-down decays,
which are absent in the \psip\ case.
The significance of this finding can be quantified by comparing the $\chi_c$ vs.\ \jpsi 
shape differences allowed by the measured data with the typical shape variations 
between the component subprocesses foreseen by theory.
The bottom panel of Fig.~\ref{fig:otherPredictions} shows rather flat shapes for the 
$\chi_c$-to-\jpsi feed-down fraction (red bands), in all considered scenarios, the
small non-flatness representing the maximal shape differences allowed by the data.
The shapes of the SDCs, instead, are significantly more different.
For example, the ratio between the \oneSzero SDC, the dominating term for $\psi$ production, 
and the \threeSone SDC, the only octet term contributing to $\chi$ production, 
changes by more than a factor of 10 from low to high \pTovM, 
as seen in the left panel of Fig.~\ref{fig:SDCratiosAndFL} (green curve).

The questions mentioned above were not addressed quantitatively in Ref.~\cite{bib:Paper1};
in the UAU scenario $\chi_c$ production is assumed, by construction, 
to be kinematically indistinguishable from $\psi$ production.
Moreover, while being a simple and natural conjecture, 
almost a mirror of the seemingly universal production scenario shown by the data, 
that model contradicts both the \mbox{$v$-scaling} rules and the HQSS relations, 
which constitute important ingredients of the NRQCD framework.
Therefore, to address those questions and to 
investigate how well the data patterns can be accounted for in NRQCD, 
we freed the $\chi_c$ cross sections from any imposed link to the $\psi$ ones 
(and between themselves), 
replacing the constraint of universal production with hypothetical $\chi_c$ polarizations. 
To evaluate the dependence of the results on this model ingredient, 
we used two options, both inspired by NRQCD calculations. 

The first variant (NRQCD\,1) assumes small and flat \chiOne and \chiTwo polarizations, 
as if, as an outcome of improved calculations, 
the unphysical and wildly varying singlet contributions 
converged to the physical and smooth \threeSone term, 
which they already closely resemble for $\pTovM > 10$ (Fig.~\ref{fig:SDCratiosAndFL}, right panel). 
The second variant (NRQCD\,2) can be considered as the state-of-the-art NRQCD scenario, 
where we inject \chiOne and \chiTwo polarizations properly determined from a previous fit 
of the \chictwooverchicone ratio, using current SDC calculations. 
Such polarizations happen to be opposite in sign and almost maximal at low \pTovM, 
potentially representing an experimental smoking-gun signature. 
This analysis leads, in both the NRQCD\,1 and NRQCD\,2 cases, 
to scenarios that are, apart from the $\chi_c$ polarizations, essentially identical to the UAU option, 
with the $\chi_{c1}$, $\chi_{c2}$ and $\psi$ mesons having approximately the same \pTovM shapes. 
We conclude that the ``universal solution" is a strong indication of the data, 
independently of specific $\chi_c$ polarization hypotheses.

Given the absence of experimental observations differentiating the production kinematics of any particular state,
the variety of \pTovM behaviours of the processes foreseen in NRQCD for the different S- and \mbox{P-wave} states,
as seen in Fig.~\ref{fig:SDCratiosAndFL}, seems excessive and unnatural.
And yet, comparing the experimental bands obtained for $\psi$, $\chi_{c1}$ and $\chi_{c2}$ cross sections 
with the corresponding ``NRQCD shapes", calculated as suitable SDC combinations, 
leads to a surprising observation. 
Despite the initial impression of unnecessary complexity, 
the NRQCD calculations can come close to reproducing the uniformity of the \pTovM trends
and the lack of polarization seen in the data, 
thanks to unexpectedly effective cancellations of heterogeneous process contributions.

The fact that NRQCD manages to mimic a scenario of almost universal and unpolarized production 
can be explained as the result of four simultaneous ``coincidences''.

1. The sum of the polarized processes contributing to $\psi$ production 
is much smaller than the unpolarized term, irrespectively of \pTovM. 
In the NRQCD language, this means that the sum $\threeSone + \threePJ$ is small 
with respect to the \oneSzero term. 
Our current results tend to indicate that the \threeSone and, especially, the \threePJ terms are individually small, 
but the present level of uncertainties prevent us from probing the idea that, as predicted by \mbox{$v$-scaling} rules, 
the \oneSzero, \threeSone and \threePJ terms have comparable magnitudes 
and that the smallness of the $\threeSone + \threePJ$ sum follows an internal cancellation,
made possible by the fact that the \threePJ and \threeSone SDCs 
have opposite signs and rather similar shapes,
for $\pTovM>5$ (Fig.~\ref{fig:SDCratiosAndFL}). 
More precise measurements of the \jpsi\ and \psip\ polarizations will 
presumably validate one of the two scenarios:
a $\lambda_\vartheta$ almost independent of \pTovM would favour the hypothesis that
the two terms cancel each other,
while a continuously changing $\lambda_\vartheta$ would suggest that both have 
small, but not identical, magnitudes.

2. While the \threeSone and \oneSzero SDCs have remarkably different slopes 
(Fig.~\ref{fig:SDCratiosAndFL}), 
adding the negative ${^3{\rm P}_{1}^{[1]}}$ SDC to the \threeSone term results in a $\chi_{c1}$ cross section shape 
almost identical to the \oneSzero term (Fig.~\ref{fig:pTovMdistrNRQCD12}, bottom left).

3. Also the $\threeSone + {^3{\rm P}_{2}^{[1]}}$ sum, constituting the observable $\chi_{c2}$ cross section, 
becomes almost identical to the \oneSzero shape (Fig.~\ref{fig:pTovMdistrNRQCD12}, bottom right). 
This happens independently of the previous observation, 
given that the ${^3{\rm P}_{1}^{[1]}}$ and ${^3{\rm P}_{2}^{[1]}}$ shapes 
(for $\pTovM<5$) are different (Fig.~\ref{fig:SDCratiosAndFL}). 
It is true that HQSS relates the ${^3{\rm P}_{1}^{[1]}}$ and ${^3{\rm P}_{2}^{[1]}}$ LDMEs,
as well as the $\threeSone \to \chi_{c1}$ and $\threeSone \to \chi_{c2}$ LDMEs, 
so that the mixing parameter $K_{\chi}$, ratio of the P- and \mbox{S-wave} LDMEs, 
is the same for $\chi_{c1}$ and $\chi_{c2}$ (Eq.~\ref{eq:chiXsectionsNRQCD}). 
At present, however, the relative proportion of the ${^3{\rm P}_{1,2}^{[1]}}$ and \threeSone terms 
is always considered as a ``free'' parameter of the theory;
the literature does not explicitly mention a relation between the 
${^3{\rm P}_{1}^{[1]}}$ and ${^3{\rm P}_{2}^{[1]}}$ SDCs,
matching the one between the LDMEs and eventually leading to
almost identical $\chi_{c1}$ and $\chi_{c2}$ observable cross-section shapes.
If such a relation existed, it would fix the mixing parameter $K_{\chi}$ and, hence, the \threeSone LDMEs,
eliminating this coincidence.

4. The $\chi_{c1}$ and $\chi_{c2}$ polarizations, 
also determined by the same ${^3{\rm P}_{1,2}^{[1]}}+\threeSone$ mixing parameter $K_{\chi}$ 
and approaching extreme physical values towards low \pTovM ($+1$ and $-3/5$, respectively), 
become almost unobservable when summed together, weighted by their respective \jpsi feed-down fractions 
(Fig.~\ref{fig:chiPolarizations}, purple band in the bottom panel). 
The surprisingly small and flat polarization resulting from this weighted sum 
would explain why no manifestations of strong polarized signals have been detected by experiments. 
Only measurements of \jpsi polarizations discriminating the two individual $\chi_c$ feed-down contributions 
from direct \jpsi production will be able to test this prediction.

Cancellations, especially when their practical realization is entrusted to the fine tuning of free parameters, 
are fragile and unstable occurrences, requiring very precise ingredients.
Further improvements in the perturbative calculations, especially for the \mbox{P-wave} SDCs, 
crucial players in this game of coincidences, are needed for more conclusive statements.

Most importantly, $\chi_{c1}$ and $\chi_{c2}$ polarization measurements have the potential 
to disclose the true heart of NRQCD, according to which different and strongly polarized individual contributions 
should be observable, at least in certain processes and/or specific phase-space corners. 
The existence of such a hidden world of diversified and polarized processes 
would be brought to light in a spectacular way by the measurement of a large difference (of order $1$) 
between the $\chi_{c2}$ and $\chi_{c1}$ polarizations 
(Fig.~\ref{fig:chiPolarizations}, orange band in the bottom panel), 
a result that should be achievable with good experimental accuracy, 
thanks to the cancellation of most systematic effects.

The opportunity provided by such a sensitive observable places NRQCD at a crossroads: 
should, instead, very similar and weak $\chi_{c1}$ and $\chi_{c2}$ polarizations be found 
(as, for example, in the UAU and NRQCD\,1 scenarios), 
the occurrence of fortunate cancellations could probably no longer be advocated.
It may still be possible that improved \mbox{P-wave} SDC calculations 
(e.g., fragmentation corrections have not yet been studied for $\chi_{c}$ singlet production) 
would rectify the $\chi_{c1}$ and $\chi_{c2}$ polarization predictions 
(e.g., in line with the NRQCD\,1 conjecture), 
bringing them close to the measured values 
and uniforming them to the semi-unpolarized context of quarkonium production. 
However, this circumstance should be seen as an even more important challenge to NRQCD: 
it would become an inescapable philosophical necessity to wonder 
whether the homogeneity of the observed kinematic patterns deserves 
a more natural theoretical explanation than a conspiracy of coincidences.

\vfill \newpage

\section*{Acknowledgements}
We are indebted to Hua-Sheng Shao,
who kindly gave us the NLO calculations of the short distance coefficients.
The work of P.F.\ is supported by FCT, Portugal,
through the grant SFRH/BPD/98595/2013,
while the work of I.K.\ is supported by FWF, Austria,
through the grant  P\,28411-N36.

\section*{References}

\bibliographystyle{elsarticle-num}
\bibliography{GlobalFit2016}{}

\end{document}